# Spin Quantum Computing with Endohedral Fullerenes
W. Harneit, Fachbereich Physik, Universität Osnabrück

## 1. Introduction

More than twenty years after its inception,[1,2] the idea of using quantum information to outperform classical computers remains an inspiring and challenging goal for the experimentalist.[3] Even in 2016, it is still unclear which of the numerous hardware concepts offers the most viable route to a large or even medium-scale implementation, i.e., a quantum processor using significantly more than a handful of quantum bits.

We take the opportunity to review here the present state of the art in using endohedral fullerenes for building a quantum computer. We will concentrate on N@$C_{60}$ and P@$C_{60}$ with their well-known favorable spin properties since they have been most studied in the quantum computing context. The related systems of metallo-fullerenes[4] and other molecular atom cages like silsesquioxanes[5] are largely disregarded here, although recent studies indicate that they may also have some favorable properties for quantum information science.[6–9]

After a brief introduction to solid-state spin quantum computing (§1), we start with fundamental considerations about the system architecture for a scalable fullerene-based quantum register element (§2), outline the main developments in endohedral fullerene materials science relevant for realizing such a register (§3), review experimental implementations of quantum operations in endohedral fullerene ensembles (§4), illustrate the ongoing quest to realize an efficient single-spin read-out for these materials (§5), and conclude with a very brief outlook on further experimental challenges and opportunities (§6).

### 1.1. Spin quantum computing

Spins can be considered as 'natural qubits' since they are often two-level systems that interact little with their environment and behave quantum mechanically even in large ensembles. Nuclear magnetic resonance on molecules dissolved in liquids (liquid NMR) allows characterizing all nuclear spins in the molecule, including their mutual spin-spin couplings, with very high precision. This facilitates the realization of even complex quantum algorithms using only molecular design.

In the early days of quantum computing (QC), NMR as a mature technique was thus at the forefront of establishing that the theoretical foundations of quantum information science were sound.[10] and it has demonstrated many of the key algorithms like prime factoring.[11] Liquid-NMR continues to drive the development of advanced QC concepts.[12]

Several limitations to the liquid-NMR approach for large-scale applications have been pointed out early on,[13] which all have their fundamental origin in the small amount of energy separating the $|0\rangle$ and $|1\rangle$ qubit states. The nuclear Zeeman splitting of a proton in an ordinary laboratory magnetic field of $B_0 = 10$ T corresponds to $\Delta E = 1.8$ µeV, which is far below the thermal energy of $k_B T = 26$ meV at room temperature. This leads (a) to small thermal spin polarization ($3\times 10^{-5}$ in the considered case; 0.5 % at $T = 2$ K) and (b) to inefficient direct detection of the low-energy electromagnetic radiation associated with Zeeman transitions, confining liquid-NMR to the use of very large ensembles ($> 10^{15}$) of identical spin systems at room temperature.

Using electron spins instead of nuclear spins can overcome some of the mentioned limitations, since the Zeeman energies are about a factor of $10^3$ higher. Electron spins also interact more strongly with the environment and are thus easier to read out. On the other hand, electrons have shorter spin lifetimes and must be carefully shielded.

### 1.2. Solid-state spin qubits

The scaling problems of the liquid-NMR approach can be overcome by going to solid-state spin systems, as was illustrated in the seminal proposal of B. E. Kane for a silicon-based computer.[14] The key ideas of Kane's paper have guided many solid-state spin proposals and it is worthwhile summarizing them here: (a) nuclear spins are realized by impurity atoms brought into a perfectly crystalline and diamagnetic matrix, ensuring minimal interaction with the environment; (b) the impurity atom simultaneously acts as electron donor and thus has an associated electron wave function including spin; (c) the electron wave function is spatially confined but can be manipulated by gate electrodes. With these ingredients, the electron spin can act as a 'gateway' or 'bus' to the nuclear spin, allowing (d) addressing, manipulating and reading out individual qubits, and (e) controlling the nearest-neighbor qubit-qubit interaction.

The single-spin read-out capability immediately offers a way to produce high spin polarization on demand, since an individual qubit can be read out repeatedly until its wave function has collapsed to a desired Eigenstate (e.g. $|0\rangle$). The interaction control, proposed originally as gate-induced spatial modulation of the electron-electron exchange interaction between neighboring qubits, represents a highly desirable element that simplifies certain quantum operations, but even 'always-on' interactions like dipolar spin-spin coupling suffice for universal quantum computing.[15]

The silicon quantum computer concept is still being actively investigated today. Being technically extremely challenging, first convincing and thus highly exciting results were obtained only in the last few years. For example, very long spin coherence times and precise quantum operations involving a Kane-type qubit were recently reported.[16–18] Two-qubit operations are however still in their infancy due to difficulties in fabricating larger quantum registers with sufficient control over each qubit.

An alternative solid-state spin qubit that has recently attracted a lot of attention is the nitrogen-vacancy center (NVC) in diamond. The NVC provides single-spin optical read-out even at room temperature.[19,20] This has lead to a fascinating and fast development of single-spin quantum control and to proposals for its use as a quantum bit.[21]

Large-scale computer architectures will require deterministic fabrication of qubits. Kane's proposal and proposals for NVC quantum computers are typical top-down approaches, which start with a macroscopic crystalline substrate and end up with almost atomic-precision fabrication of spin qubits (dopant atoms in silicon or color centers in diamond), which is hard to achieve by ion implantation. This is even more challenging for the NV center in diamond since (a) fabrication technologies for diamond are less mature than for silicon and (b) the formation of NVC color centers is a two-step process requiring (i) N ion implantation and vacancy creation, and (ii) recombination of both defects to produce a NV center.[22]

The single-spin read-out capability of the NVC however makes it an excellent read-out system for other spins, including qubits in spin-based QC proposals. Already, advances in quantum control of singly addressed NVCs have allowed to detect and control larger and larger portions of the 'dark' spin environment, ending up with current trends to map out the 3D electron spin environment using a single NVC[23] and to realize even single-molecule NMR based on NVC detection.[24]

### 1.3. Paramagnetic endohedral fullerenes

In this review we will concentrate on N@$C_{60}$ and P@$C_{60}$, which we call *paramagnetic* or *pnictide* endohedral fullerenes (PEFs). In the quantum computing context, PEFs are officially classified as 'impurity spin' systems in similarity with Kane's Si:P proposal or with the alternative nitrogen-vacancy center (NVC) in diamond (see §1.2).

In contrast to solid-state spin qubits based on dopant atoms or color centers, molecular spin systems provide a bottom-up route to large-scale quantum register fabrication.[25,26] In a first stage, identical molecular qubits can be fabricated (incidentally also by ion implantation for PEFs) and purified chemically, and only later assembled in almost arbitrarily complex architectures using chemical engineering approaches.[27] This decoupling of qubit creation and assembly is a potentially big advantage for device architecture and scalability and one of the main drivers for using PEFs as a QC resource.

## 2. System Architecture and Computing Models

We review basic approaches on how the central processing unit of a PEF-based quantum computer could be constructed. We start (§2.1) by identifying possible qubits in the multi-level spin system and point out some important consequences of the high-spin character of the electron spin. In §2.2, we discuss the nature of spin-spin interactions in PEFs and the resulting central architecture of a one-dimensional quantum register.

### 2.1. Qubits in the Multi-Level Electron-Nuclear Spin System

Pnictide endohedral fullerenes consist of a fullerene cage with a paramagnetic atom of IUPAC group 15 (pnictides), which is freely suspended in the cage center by dispersive van-der-Waals forces.[28] This constitutes a highly shielded symmetric spin system with long coherence times, favorable for spin quantum computing. The fullerene cage acts as a robust handle to the fragile atom, conferring thermodynamic and chemical stability and hence some flexibility in designing devices.

In N@$C_{60}$ (Refs. 29,30) and P@$C_{60}$ (Refs. 31,32), the encapsulated atom has a half-filled *p* shell and hence a total electron spin $S = 3/2$ with Zeeman sublevels characterized by the quantum number $m_S = \pm 1/2, \pm 3/2$. Pnictide atoms also have a nuclear spin $I = 1/2$ ($^{15}$N, $^{31}$P) or $I = 1$ ($^{14}$N) with corresponding quantum number $m_I$. The effective spin Hamiltonian of an isolated PEF molecule is given to first order by

$$H_0(B_l)/h = \nu_S(B_l)\, S_z - \nu_I(B_l)\, I_z + A\, S_z I_z + D\, (S_z^2 - 5/4), \qquad (1)$$

where $\nu_{S,I}(B_l) = g_{e,n}\, \beta_{e,n} B_l/(2\pi)$ are the Larmor frequencies at effective local magnetic field $B_l$, $g_{e,n}$ are the isotropic g-factors and $\beta_{e,n}$ the magnetons for electron and nucleus, respectively. Note that g-factors, and hence Larmor frequencies, are signed quantities. The scalars $A$ and $D$ are frequencies characterizing isotropic hyperfine interaction *(hfi)* between electron and nuclear spin and axial zero-field interaction *(zfi)*, which is characteristic for high-spin systems and originates from dipolar interaction between the three constituent electrons.[25] Parameters for the most important PEF molecules are summarized in Table 1.

Electron and nuclear spins can be discussed as independent resources; they have well-defined quantum numbers that can be used to characterize a PEF spin eigenstate as $|m_S, m_I\rangle$. Spin control is achieved by driving allowed transitions, either in the microwave (MW) range for electron spin resonance (ESR, $\Delta m_S = \pm 1, \Delta m_I = 0$) or in the radio-frequency (RF) range for nuclear magnetic resonance (NMR, $\Delta m_S = 0, \Delta m_I = \pm 1$).

**Table 1. Typical parameters for the PEF spin Hamiltonian (Eq. 1).** *The Larmor frequencies $\nu_{S,I}$ depend on the effective local magnetic field $B_l$ while D depends on the local symmetry (see §3.3); $D = 0$ for 'bare' $C_{60}$ in an isotropic environment. All values are given in MHz.*

| quantity | multiplier | $^{14}N@C_{60}$ | $^{15}N@C_{60}$ | $^{31}P@C_{60}$ |
|---|---|---|---|---|
| $\nu_S$ | $\times(B_l/T)$ | ⊢ – – – – 28000 – – – – ⊣ | | |
| $\nu_I$ | $\times(B_l/T)$ | 3.08 | −4.32 | 17.25 |
| $A$ | | 15.8 | 22.2 | 138.5 |
| $D$ | | 0—30 | 0—30 | 0—127 |

First-order hyperfine coupling ($A$) splits the electron spin resonance ($|\Delta m_S| = 1$) into as many equidistant lines as there are nuclear spin states ($2I + 1 = 2$ or $3$). These lines are still triply degenerate in the absence of zero-field interaction ($D = 0$). The degeneracy is lifted for $D \neq 0$, leading to $(2I + 1) \times (2S)$ selectively addressable ESR lines (see Fig. 1). Similarly, there are $(2I) \times (2S + 1)$ NMR resonances with a first-order splitting in $m_S$ and a first-order degeneracy in $m_I$ for the $I = 1$ system $^{14}N@C_{60}$, lifted only by the second-order hyperfine interaction. Second-order *hfi* can lead to other effects in high-resolution spectroscopy (see §4.1 for its use in qubit characterization) but is otherwise less relevant for quantum computing due to its small size.

A total number of eight ($^{15}N$, $^{31}P$) or twelve ($^{14}N$) Zeeman sublevels result from the static Hamiltonian (see Fig. 1). Up to three qubits ($2^3 = 8$) may thus be defined and entangled,[33] and even four qubits would be possible in the $S = 3/2$, $I = 3/2$ spin system of the (not yet experimentally reported) $As@C_{60}$ analogue.[34] The four levels of the high $S = 3/2$ electron spin can in principle be used to encode either two qubits[33,35] or a *ququart* (a four-level generalization of the two-level qubit), which is useful in some quantum computing schemes and has been studied for analogous $I = 3/2$ nuclear spins in the NMR domain.[36,37] In both cases, however, transition-selective addressing is mandatory, requiring a finite 'quadrupolar' splitting Hamiltonian of the form $QI_Z^2$. In the PEF Hamiltonian, this corresponds to non-zero $D$ in the *zfi*-term, which leads to a lifting of the ESR transition degeneracy (see Fig. 1).

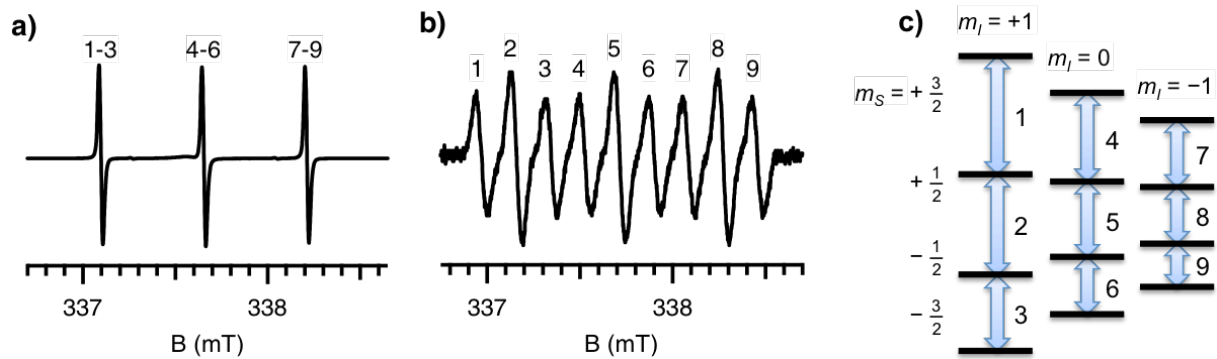

**Fig. 1. Electron spin resonance (ESR) spectra of $^{14}N@C_{60}$ diluted in $C_{60}$.** *(a) Polycrystalline powder sample; (b) sample in anisotropic crystalline host matrix (BrPOT); (c) energy levels and allowed ESR transitions. Adapted from Ref. 38, with permission from Wiley & Sons.*

The *zfi* term is intimately linked to the shape of the total electron wave function of the encapsulated atom, which is in turn strongly influenced by the fullerene cage. It is absent for *effective* high cage symmetries such as occur for low-symmetry fullerenes dissolved in liquids due to rapid tumbling or for 'bare' (i.e., chemically unmodified) $C_{60}$ cages even in a crystalline environment at room temperature, where the molecules rotate freely. Ways to change *zfi* size and orientation will be discussed in more detail in §3.3. Here, we note that the presence or absence of *zfi* has notable consequences on the way that all four sublevels of the electron spin can be addressed and hence used for quantum information.

The nuclear spin can be seen as an independent qubit, which is even better shielded from the environment than the electron spin. The nuclear spin has much longer relaxation and coherence times, but manipulating it with resonant RF pulses is also much slower. §4 will show that there is a net gain for quantum computation; the gain in coherence time is larger than the loss in control speed. Therefore, the nuclear spin has been considered as the primary logical qubit in several theoretical proposals, with the electron spin serving as a 'gateway' or 'bus' only.[39–41]

## 2.2. Spin-Spin Interactions and the Linear Qubit Register

Experimental studies on the concentration dependence of ESR line broadening[42] and theoretical calculations[30] show that the exchange interaction between two endohedrals is negligible, leaving magnetic dipole interaction as the only strong coupling. For two adjacent endohedrals at a typical[43] center-to-center distance of 1.0 nm the magnetic dipole interaction between electron spins can be as large as ~ 50 MHz. This dipolar coupling between electron spins scales as $r^{-3}$ with the fullerene distance, is anisotropic, and is not confined to nearest neighbours since it acts 'through space'. Any disorder in the spatial arrangement of fullerenes may therefore be a considerable source of decoherence.

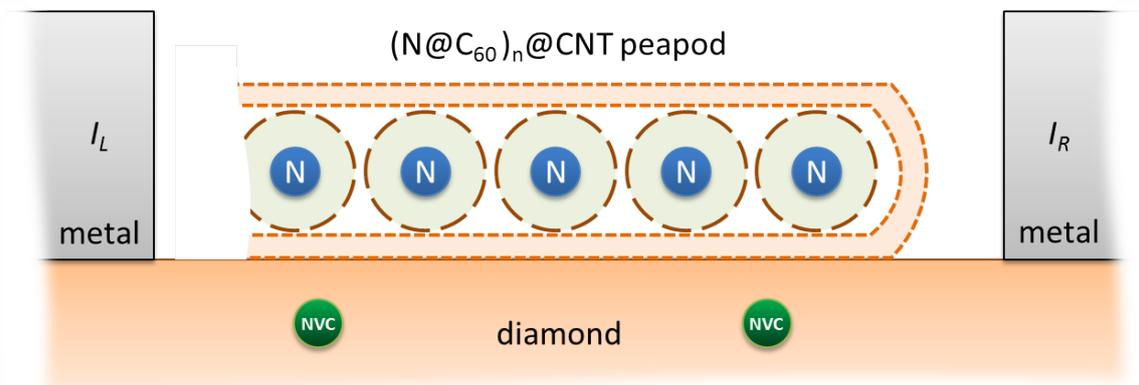

***Fig. 2. Sketch of a linear spin-chain quantum register with optical readout.*** *N@$C_{60}$ or P@$C_{60}$ electron-nuclear spin qubits are lined up inside a carbon nanotube (CNT), which is placed on top of a diamond substrate with shallow nitrogen-vacancy centers (NVC). A current is driven through nearby metal electrodes, producing a magnetic field gradient that enables addressing of individual fullerene or NVC spins with MW or RF pulses. Thus, controlled quantum operations between fullerene spins coupled by magnetic dipole interaction can be performed. Initial and final spin states are prepared and read out optically via spin-spin coupling to the NVC readout qubits.*

This suggests that a one-dimensional arrangement of spins in a *linear qubit register* will provide the highest degree of interaction control, and this arrangement has been at the focus of theoretical proposals for endohedral fullerenes.[25,26,35,44,39,40,45,41,46] A linear chain also provides a simple means to make every qubit addressable when a magnetic field gradient is applied along the chain direction.[39] Taking *z* as the direction of the external magnetic field $B_0$, as the chain direction, and as the direction of the (constant) gradient $\nabla B_G$, the local magnetic field $B_l(z) = B_0 + z\,\nabla B_G$ changes as a function of $z$ and hence so do the resonance frequencies $\nu_{S,I}$. This leads to the total spin Hamiltonian of a linear chain of $N$ identical PEF molecules

$$H_N = \sum_{j \leq N} H_0\left(B_l(z_j)\right) + h \sum_{j<k} D_{jk}(|z_j - z_k|)\, S_{zj}\, S_{zk}, \qquad (2)$$

where $D_{jk}(r) = (51.9\ \text{MHz}) \times (r/\text{nm})^{-3}$ and $H_0(B)$ is given by Eq. (1) with $B_l(z) = B_0(1 + \Delta(z - z_0))$ and $\Delta = (\nabla B_G)/B_0$.

At first sight, the always-on nature of the dipolar coupling between electon spins is an undesirable feature for quantum computing. Dipolar interaction in the chain extends to more distant fullerenes, leading to a highly structured ESR spectrum for each qubit in the chain; individual addressing is nevertheless still possible if the electron spins are normally polarized (see Fig. 3). Furthermore, dipolar coupling continually drives a state-dependent evolution of the electron spins, making them useless as memory qubits or requiring continuous active decoupling by microwave pulses.[15]

The solution is to take the nuclear spins as primary logical qubits. The combination of an electron-nuclear spin system at each fullerene site then turns out to provide an elegant way to control qubit-qubit interaction.[40,45,41] Direct dipolar coupling between nuclear spins of adjacent fullerenes is negligible ($< 10$ kHz) and may be selectively enhanced by using the electron spin as a 'bus' or 'auxiliary' qubit. Several strategies have been proposed[35,45,47] for 'swapping' quantum information between the logical nuclear spins and the bus electron spins that are otherwise kept in a polarized ground state. Logical two-qubit operations then proceed by swapping out nuclear qubit states onto the electron spins, engineering electron-spin-only quantum operations, and 'swapping back' the information onto the nuclei.

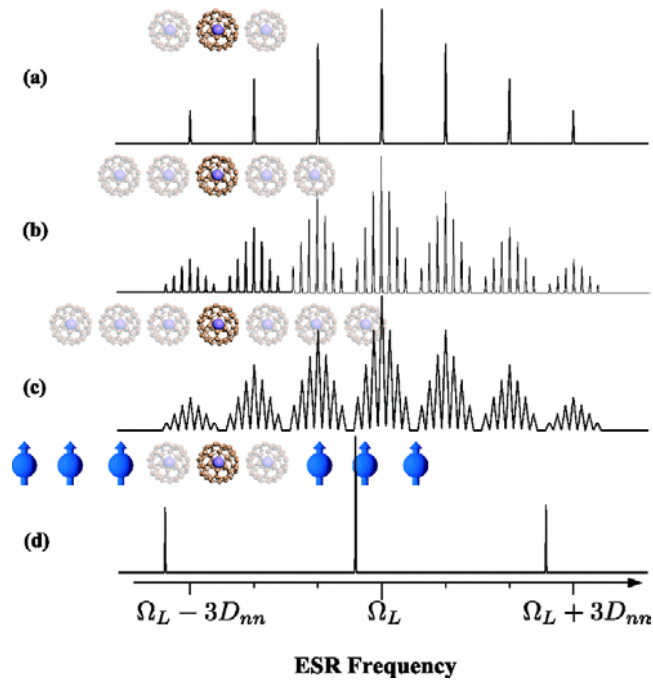

***Fig. 3. PEF electron spin transition frequencies as a function of chain length. (a)*** *Nearest neighbors only;* ***(b)*** *two and* ***(c)*** *three neighbors on each side;* ***(d)*** *infinite chain where all but the nearest neighbors are polarized (i.e., in the 'passive' ground state). Adapted from Ref.* [45], *with permission from The American Physical Society.*

Even with traditional approaches to create an electron-nuclear swap operation, which takes up most of the gating time, thousands of fullerene qubits may be made to function in a single quantum register.[41] Furthermore, Morton *et al.* have shown[48] that nuclear phase gates may be realized much more rapidly (see §4.2), potentially increasing this number. Finally, there have been proposals to build a quantum network out of several independent linear chains by entangling them at a distance.[45,49]

## 3. Materials Science Aspects

Having identified the linear spin chain as the basic architecture for most QC concepts, we now discuss materials science aspects specific to the QC application of endohedral fullerenes. Broader reviews of PEF materials science can be found in several reviews; we recommend a book chapter by B. Pietzak *et al.* on early work[50] and the chapter of K. Porfyrakis in this book[51] for more recent work.

### 3.1. Synthesis, Purification, Stability

The standard synthesis of PEF molecules by ion implantation (present efficiency limit: $3 \times 10^{-4}$) and subsequent multi-step HPLC purification is quite precise and affords up to 100% purity;[52] yet it is extremely laborious. The low overall synthesis yield still constitutes one of the major obstacles in quantum computing research. A significant observation is that chemical derivatives (or 'adducts') of $N@C_{60}$ can be separated from their 'empty' $C_{60}$ counterparts using the same HPLC methods.[53] This allows using larger sample amounts during a possibly low-yield chemical adduct synthesis while enriching the endohedral fraction *of the product* during a post-synthesis HPLC workup step.

The thermal, chemical, and optical stability of PEF complexes have been investigated several times.[54–57] Thermal decomposition sets in above $130 - 150$ °C; the thermal stability is thus sufficient for safe handling at room temperature and even allows chemical reactions to occur under reflux in toluene. It can be increased by encapsulation into carbon nanotubes.[55] The situation is less clear for the optical stability; recent studies indicate that while properly cooled PEF molecules are not harmed by high-power optical irradiation, insufficient cooling may induce thermal instability.[56] This may have caused confusion in some optical stability investigations.

### 3.2. Integration in Solid-State Architectures

A solid-state quantum computer needs complex and precise spatial arrangements of fullerenes. The moderate thermal stability of PEF molecules implies that one cannot use vacuum sublimation for thin-film deposition or other high-temperature processes, such as crystal growth or encapsulation into carbon nanotubes driven by thermal diffusion. Hence alternative 'cold' deposition[38,58,59] and encapsulation[60,61] methods have been developed for PEFs, with potential applications for other thermo-labile molecules.

Given these preparation methods, self-assembly of fullerenes is then possible as has been investigated in numerous studies. Here, we arbitrarily quote recent (sublimation-based) studies of self-assembly on surfaces, where even 2D architectures are possible.[62,63] In 3D matrices, ordering may include chain-like or at least one-dimensional motifs such as in liquid crystals[64–66] or in pore-forming organic compounds.[38,67,68]

Finally, the possibly best defined and most highly ordered linear arrangement proceeds by encapsulation into carbon nanotubes (CNTs).[60,61] The electronic and optical properties of CNTs make this particularly attractive as it could permit electrical readout[26,45] or distant entanglement schemes in hybrid architectures.[49,69,70]

### 3.3. Control of the Zero-Field Splitting

The zero-field interaction (*zfi*) is essential for several QC proposals (see §2) since it allows transition-selective addressing of all spin sublevels. It is important to note that the *zfi* is anisotropic and reflects the molecular symmetry, hence not only the size of $D$ has to be controlled but also its orientation.

For highly symmetric systems (spherical or cubic), $D$ is exactly zero. The same holds for fast orientational averaging, as occurs in liquid samples and even in polycrystalline $C_{60}$ samples at room temperature. Small $D$ values are observable at lower temperatures in solid PEF samples: N@$C_{60}$-doped $C_{60}$ powder,[71] P@$C_{60}$-doped powder,[72] N@$C_{60}$-doped $C_{60}$ crystals.[73] The $D$ values are distributed in both amplitude and orientation, resulting in featureless and broad ESR lines. Nevertheless, valuable spectroscopic information can be gleaned even in this situation using the *nutation* behavior specific to $S = 3/2$ systems (see §4.2).

Larger $D$ values can be engineered using either covalent or non-covalent chemical functionalization. Covalent modifications have been investigated since the early days: Dietel *et al.* have shown the first mono-adducts characterized by $D$ values of around 10 MHz, and regio-selective attachment of up to six addends of N@$C_{60}$, affording a nearly cubic hexakis adduct.[74] Franco *et al.* have systematically varied addends in mono-adducts and investigated their effect on $D$.[75] First fullerene dimers were reported by Goedde *et al.*[76] and many contributions were made by the Porfyrakis group.[53,57,77,78] Covalent functionalization leads to $D$ values of up to 15 MHz for N@$C_{60}$ (comparable to the hyperfine interaction), but the resulting EPR spectra are usually broad and of little use for quantum information due to random orientation of the molecular axes induced by the covalent addends. This can in principle be avoided if stiff addends are used and if the adducts are grafted or self-assembled on a geometrically ordering substrate such as a flat surface or a carbon nanotube.

Non-covalent interactions of fullerenes in molecular environments provide an alternate route towards useful zero-field effects.[27] The basic idea here is that the pseudo-aromatic fullerene cage may act as a transducer for symmetry-lowering $\pi - \pi$ interactions with the environment. This is already obvious from the finite $D$ values observed[71–73] in polycrystalline samples of 'bare' (unmodified) endohedral fullerenes (vide supra), which is due to $C_{60}$-$C_{60}$ interaction. If the environment is highly ordered, this may lead to induced order of the fullerenes.

Several groups investigated interactions with molecular recognition systems that wrap around the fullerene.[75,79,80] While leading to zero-field interaction $D$ of similar size as due to covalent $C_{60}$ chemistry, these systems are also difficult to orient. Surprisingly, partial orientation of the near-spherical $C_{60}$ fullerenes was found in liquid crystals,[64,65] but the $D$ values were rather small and significant residual disorder was found.

Inclusion of endohedrals in pore-forming organic crystals (halo-phenoxy-triazines like Br-POT[81]) has shown the best *zfi* control so far.[67,82] Inclusion of fullerenes in *cylindrical* Br-POT pores leads to partial orientation with significant disorder due to co-included solvent molecules and is thus similar to the liquid crystal case. Somewhat surprisingly, a second mode of co-crystallization was found that leads to the formation of a highly oriented arrangement of individual host-guest 'pockets' for fullerenes. This structure induces near-perfect rotational order for the fullerenes as well as considerable $D$ values exceeding those obtained by covalent functionalization. This system was therefore used to realize two-qubit gates (see §4.3).

Inducing orientational order is also vital for investigations of PEF dimers containing pnictide spins in both monomers because it would allow control over the dipolar spin-spin interaction, which is necessary for realizing controlled two-qubit interactions (see

§4.4). Similar orientation approaches using other matrices such as zeolites or metal-organic frameworks have yet to be explored systematically.

## 4. Quantum Operations and Entanglement

We give a short overview over key experiments on quantum operations in magnetically dilute ensembles of group-V endohedrals, which illustrate some of the theoretical proposals outlined in §2. Starting with spin-relaxation and coherence times, we review 1-qubit operations for electron and nuclear spin, and intra-fullerene electron-nuclear 2-qubit operations such as coherent state transfer and entanglement. We end with a discussion of the status for inter-fullerene two-qubit operations.

### 4.1. Spin Relaxation and Coherence

In the coupled electron-nuclear spin system of PEFs, the electron spin–lattice relaxation time $T_{1e}$ is of fundamental interest. $T_{1e}$ dominates electron spin coherence $T_{2e}$ at room temperature[83] and nuclear spin coherence $T_{2n}$ at low temperature.[84] The mechanisms inducing spontaneous spin flips and hence $T_{1e}$ relaxation are quite sensitive to the environment of the fullerenes.[85]

In liquids, collisional processes may distort the fullerene and hence modulate terms in its spin Hamiltonian, most notably the zero-field interaction (*zfi*). Indeed, chemically modified N@$C_{60}$ molecules with a permanent *zfi* show accelerated relaxation in liquids, whereas *g*-factor and hyperfine interaction are quite insensitive to symmetry-lowering distortions of the fullerene cage.[74] The *zfi*-based collisional relaxation model was also invoked to explain the faster relaxation[26] in the P@$C_{60}$ molecule, which also shows[67,72] larger matrix-induced *zfi* than N@$C_{60}$ does.[71,82]

As an alternative to the collisional $T_{1e}$ model, an Orbach process was proposed[83,85] to explain relaxation above $T > 150$ K, essentially based on the temperature dependence of $T_{1e}$ and $T_{2e}$ in frozen solution and powder samples. The Orbach process describes relaxation via low-lying excited states outside the usual spin ground state manifold. These states were identified with molecular vibrations. Differences of $T_{1e}(T)$ in frozen solutions of toluene and CS$_2$ were attributed to the excitation of different molecular vibrations of the $C_{60}$ cage,[85] but direct evidence for the suppression of the lowest-energy H$_g$(1) 'squashing' mode of $C_{60}$ in frozen toluene solutions is lacking. Furthermore, the H$_g$(1) mode[86] is clearly symmetry-breaking for the confinement potential of the trapped atom (thus capable of inducing a transient *zfi*), whereas the A$_g$(1) "breathing" mode (invoked to explain the toluene data[85]) could only modulate the isotropic hyperfine interaction, which is not sufficient to establish thermodynamic equilibrium in the spin system.[87] Therefore, the direct identification of Orbach energies with molecular vibration energies remains questionable. Finally, there must be further mechanisms determining $T_{1e}$ below about $T = 150$ K, which remain to be identified.

Electron spin dephasing $T_{2e}$ and relaxation $T_{1e}$ have the same temperature dependence near room temperature. There is a systematic difference between 'inner' ($m_S = 1/2 \leftrightarrow -1/2$) and 'outer' ($\pm 1/2 \leftrightarrow \pm 3/2$) transitions, $T_{2e}^o = (2/3)\, T_{2e}^i = (4/9)\, T_{1e}$,[83] which could be measured independently due to a rare echo modulation effect occurring in $^{14}$N@$C_{60}$.[88] At temperatures lower than about $T = 100$ K, $T_{2e}$ is independent of temperature in powder samples while $T_{1e}$ still rapidly grows. Several studies[84,85] show that $T_{2e}^o$ may be much shorter than $T_{2e}^i$ for N@$C_{60}$, but interestingly this is not the case for P@$C_{60}$ with its much larger zero-field and hyperfine interactions.[68,89,90]

At low temperatures, the dominant $T_{2e}$ dephasing mechanism can be attributed to uncontrolled dipolar interaction between different endohedral spins.[84,90] This can be remedied by achieving better coupling control, as should be possible in a linear spin chain (see §2.2). Preliminary studies of P@$C_{60}$ at high endohedral spin dilution show that dephasing by impurity nuclear spins then takes over, leading to a lower limit of the achievable $T_{2e} \geq 3.9$ ms at $T = 5$ K.[90]

The nuclear relaxation times are much harder to access, but specialized experiments provided the estimates $T_{2n} = 135$ ms (measured at $T = 10$ K using a coherence transfer method[84]) and $T_{1n} \approx 12$ hours (measured at $T = 4.2$ K in a dynamic nuclear polarization experiment[91]). The 'best' PEF relaxation data published to date for $4.2 \text{ K} \leq T \leq 10 \text{ K}$ and $B_0 \approx 0.35$ T are gathered in Table 2. These data show that there is room for improvement in materials purification since they do not meet the standard assumption for clean material, which is that $T_{2e}$ and $T_{2n}$ should reach $T_{1e}$ at low temperatures.

**Table 2. Highest reported spin relaxation times for PEF powder samples at $T = 4.2 - 10$ K.**
*$T_1$: spin-lattice relaxation, $T_2$: spin-spin relaxation (coherence) for electron (e) and nuclear (n) spin.*

|           | $T_{1n}$ | $T_{2n}$ | $T_{1e}$ | $T_{2e}$ |
|-----------|----------|----------|----------|----------|
| time / ms | 4×10⁷    | 135      | 1300     | 3.9      |
| PEF type  | $^{14}$N@$C_{60}$ | $^{15}$N@$C_{60}$ | $^{31}$P@$C_{60}$ | $^{31}$P@$C_{60}$ |
| reference | 91       | 84       | 89       | 90       |

### 4.2. Qubit Addressing and Single-Quantum Gates

Single-qubit rotations can be achieved by microwave (MW) irradiation for the electron spin or by radio-frequency (RF) irradiation for the nuclear spin. The electron spin nutation behavior is more complicated for the $S = 3/2$ spin system than for $S = 1/2$ spins.[72] In general, $S = 3/2$ spin operators have to be used for a complete description. For vanishing zero-field interaction, the isolated electron spin 'classically' behaves exactly as a $S = 1/2$ system, e.g., a π pulse of the same length exactly inverts the sub-level populations for $S = 1/2$ or $S = 3/2$. For superpositions or coherences, such as those produced by π/2 pulses, one has to distinguish between 'inner' (involving only $|m_S| = 1/2$ levels) and 'outer' (involving $|m_S| = 1/2$ and 3/2 levels) coherences.[84] Both will be excited for turning angles other than multiples of π.

If there is interaction with other spins, e.g. hyperfine interaction with the nucleus or dipolar interaction with neighboring electron spins, the different coherences lead to new effects due to conditional phase evolution. For the hyperfine case, for example, the phase evolution will depend on both quantum numbers, $m_S$ and $m_I$.[88] Such effects can in principle be used to engineer desired phase evolutions across a spin chain, but the full theory quickly becomes opaque. Therefore, the standard operating model for a fullerene-based linear-chain quantum register is to use only π pulses on the electron spins and avoiding electron-spin coherences as much as possible. [40,45,41]

The situation is conceptually simpler if there is a finite *zfi*, since transition-selective ESR and NMR pulses then give access to the full Hilbert space. This means that qubits can be arbitrarily defined, and that single MW or RF pulses can be treated as if they applied to a 'fictitious' $S = 1/2$ spin. Examples for using this addressing mode can be found in the works of C. Meyer *et al.*[72] and of Naydenov *et al.*[68]

In practical experiments, electron-spin rotations with non-selective MW excitation can suffer from a number of 'imperfections' leading to gate errors. Morton *et al.*, have used PEFs to study ways to quantify such errors[93] and to suppress them by applying methods known from NMR, such as BB1 composite pulses. [94]

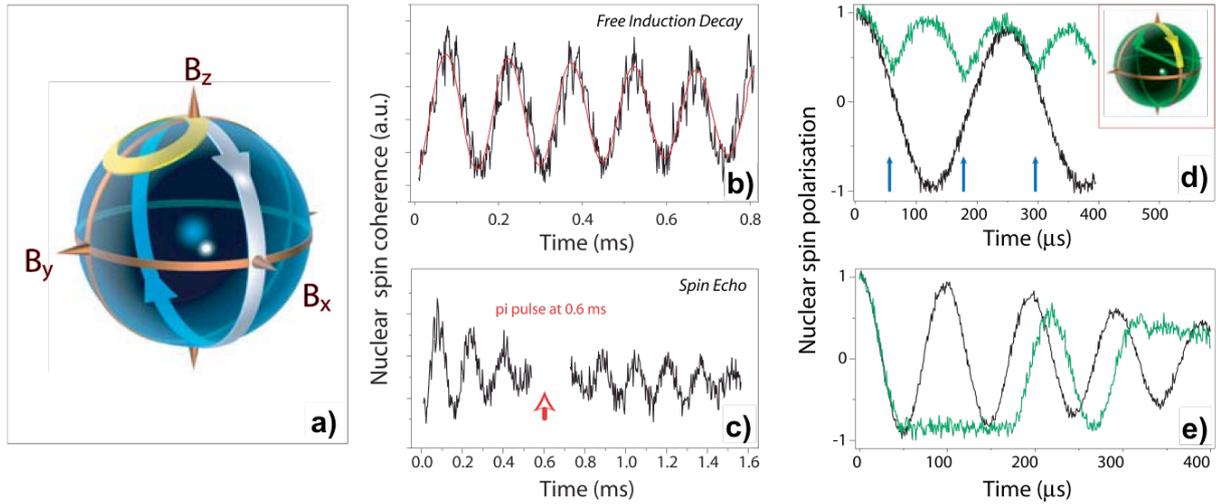

*Fig. 4. Fast nuclear phase gate using a driven electron spin.* **(a)** *Effect of applying resonant (white) or detuned (yellow) microwave pulses on the electron spin magnetization.* **(b)** *Free induction decay, and* **(c)** *spin echo of the nuclear spin indicate that the nuclear coherence time $T_{2n}$ is limited only by the electron spin $T_{1e}$.* **(d, e)** *Black: unperturbed Rabi oscillations between the two nuclear spin states of the $^{15}$N atom; green: nuclear spin evolution under the influence of bang-bang decoupling pulses. Adapted from Ref.* [92], *with permission from Wiley & Sons.*

Since the nuclear spin has far superior relaxation and coherence times (see §4.1), it is considered the primary 'logic' qubit in the endohedral spin system.[39–41] RF pulses to manipulate it are however far slower than the MW pulses used to manipulate the electron spin, potentially limiting the number of logical operations. Morton *et al.* have shown[48,92] that in the coupled spin system the electron can be used as an auxiliary resource to dramatically speed up nuclear spin manipulation.[95] In essence, driving an electron spin transition connected to the nuclear spin can create arbitrary changes of the nuclear phase with high precision, but at the time scale of the electron spin Rabi frequency ('bang-bang' pulse duration ~100 ns, see Fig. 4). The resulting fast nuclear phase gates may help to make quantum computing with endohedrals universal in the sense that a sufficient number of logical operations can be performed with integrated error correction.[96] First work demonstrating how to integrate error correction with quantum operations on endohedral fullerene qubits can be found in the coherence transfer experiment discussed in the next section.[84]

### 4.3. Electron-Nuclear 2 Qubit Operations

The spin structure of PEFs has been used to explore the feasibility of entanglement and coherence transfer between electron spins and nuclear spins. Most experiments were performed on $^{15}$N@$C_{60}$ molecules diluted in a polycrystalline $C_{60}$ matrix and hence used non-selective addressing.[48,84,97,98] One study[68] used transition-selective addressing of magnetically dilute P@$C_{60}$:$C_{60}$ encapsulated in a molecular matrix,[67] which provides a high degree of orientational order and hence a clear *zfi* structure.

In the work of Mehring *et al.*,[97,98] the 'outer' electron spin states of $^{15}$N with $|m_S| = 3/2$ were used while the 'inner' $|\pm 1/2\rangle$ states were disregarded (see Fig. 5). Electron-nuclear pseudo-entanglement was demonstrated and quantified via quantum state tomography. Very high fidelities ($F > 0.99$) for state preparation were found in these experiments, both for pseudo-pure and pseudo-entangled states.[98] The prefix 'pseudo' indicates that at the relatively high temperature of $T = 50$ K (chosen for the experimental reason of otherwise too large $T_{1e}$ values), the density matrix is still separable and hence 'pseudo-entangled'. A calculation shows that real entanglement can be created below about $T = 7.5$ K at W-band conditions ($B_0 = 3.5$ T, $f_{MW} = 95$ GHz), which is realistically achievable.[97]

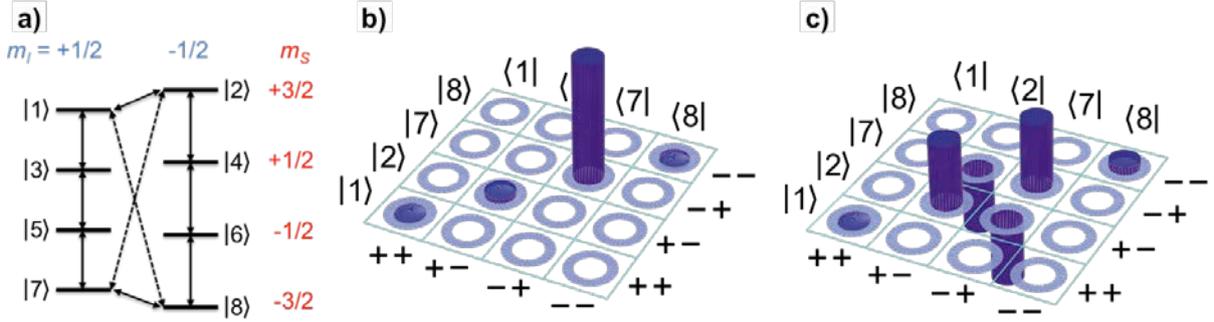

**Fig. 5. Electron-nuclear pseudo-entanglement.** *(a) Energy levels of $^{15}$N@$C_{60}$ and (b, c) quantum state tomography. (b) Experimentally determined density matrices of the pseudo-pure state $|7\rangle = |m_S = -3/2, m_I = +1/2\rangle \equiv |-+\rangle$, $\rho_7 = |7\rangle\langle 7|$, and (c) of the pseudo-entangled Bell state $\Psi^-_{27} = (|2\rangle - |7\rangle)/\sqrt{2} \equiv (|+-\rangle - |-+\rangle)/\sqrt{2}$. Adapted from Ref. [97], with permission from The American Physical Society.*

Despite the good fidelity of state preparation, the lifetime of the entangled state turned out to be limited to ~ 0.2 µs while the single-quantum coherence time was much higher ($T_{2e}$ ~ 3 µs). In a follow-up experiment,[68] selective addressing was used to explore entanglement between two arbitrarily defined qubits in P@$C_{60}$ (one electron spin level pair with $m_I = 1/2$ and $m_S = \pm 1/2$ and the other with $m_I = -1/2$ and $m_S = -1/2, -3/2$). State preparation was again found to be of very high fidelity, but the resulting lifetime of the entangled state was even shorter. Both papers attributed entangled-state decoherence to unresolved *zfi*, but detailed investigations in this direction have yet to be carried out, e.g., by comparing with entanglement involving only $m_S = \pm 1/2$ states, or using 'quadrupolar' quantum state tomography.[99]

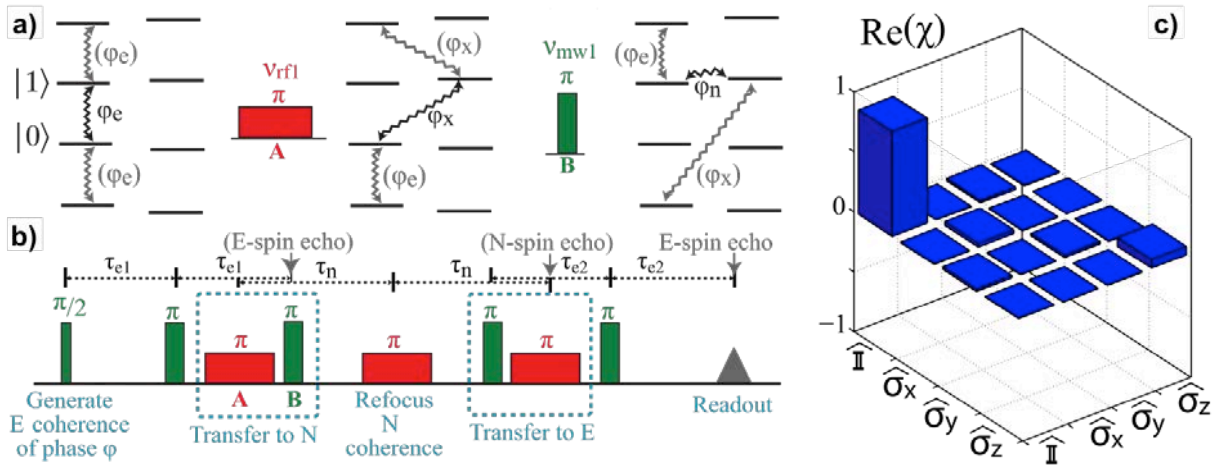

**Fig. 6. Transfer of a qubit state from the electron spin to the nuclear spin of $^{15}$N@$C_{60}$** *(diluted in $C_{60}$) within the $m_S = \pm 1/2$ subspace. (a) Coherences are depicted by wiggly lines; 'unwanted' coherences generated by the initial $\pi/2$ pulse on the $S = 3/2$ electron spin are shown in gray. (b) The full two-way transfer sequence. (c) Quantum process tomography matrix $\chi$ for the two-way transfer, yielding a fidelity $F = 0.88$. Adapted from Ref. [84], with permission from The American Physical Society.*

An essential ingredient of the linear chain register (§ 2.2) is to 'swap' spin quantum information between 'logical' nuclear and 'bus' electron spin qubits. An experiment demonstrating this type of coherence transfer was realized in dilute $^{15}$N@C$_{60}$ powder by Brown *et al.*[84] This experiment operated only on $m_S = \pm 1/2$ states and disregarded the $|\pm 3/2\rangle$ states due to their poor coherence properties. The experiment itself used a straightforward transfer sequence from electron spin coherence to nuclear spin coherence, dispensing with pseudo-pure state creation. On the other hand, a judicious choice of refocusing pulses was interspersed to maximize gate fidelity. A two-way (back-and-forth) gate fidelity of $F = 0.89$ was obtained that could be increased to $F = 0.93$ using the previously optimized[94] BB1 electron spin operation pulses.

### 4.4. Towards Electron-Electron 2 Qubit Operations

Studies of two-qubit operations or entanglement between electron spins localized on neighboring endohedral fullerenes are still lacking. In the first proposals,[25,27] chemical dimers were invoked as a likely route towards such studies, but this approach has proved to be harder than anticipated. The main reason is the low overall synthesis yield of endohedral fullerenes, which makes chemical coupling reactions hard to realize.

Although C$_{60}$ dimers can be easily produced and 'half-endohedral' dimers have been demonstrated using various approaches,[76,77] all coupling reactions bear the inherent danger of spin loss. The Porfyrakis group has studied this problem in some detail; recent achievements include nearly lossless Bingel reactions[57] as well as more sophisticated dimer coupling schemes[77] that take into account the limitations of low-yield endohedral synthesis. A doubly-endohedral dimer was already reported[78] but its spin properties could only be investigated in a disordered solid-state sample. Since the Hilbert space of a ($^{14}$N@C$_{60}$)$_2$ dimer has dimension 144, highly ordered samples (see §3.3) are needed for further progress.

An alternative route towards controlled fullerene-fullerene coupling is given by non-covalent encapsulation in matrices such as carbon nanotubes (see §3.2), with the added benefit of a natural linear orientation.[60,61] To investigate such samples however requires that a single-spin read-out be found.

### 5. The Single-Qubit Read-Out Challenge

There is a great fundamental and practical need for finding single-spin readout methods compatible with endohedral fullerenes. This will allow using solid-state architectures based on individual spin chains rather than big ensembles of fullerenes or fullerene dimers. Spin chains (see §2.2) bring advantages in interaction control, desirable both for reduced error rates (suppression of uncontrolled dipolar interactions, see §4.1) and for making quantum-computing universal through local interaction gating. Even quantum cellular automaton (QCA) architectures with only global interaction control[35] would benefit. Moreover, a strong (projective) read-out allows for efficient qubit state preparation and polarization, enabling more elaborate quantum protocols. In the following, we first review the limits of traditional 'inductive' microwave detection, which has recently progressed almost to single-spin detection. The remaining sections describe indirect electrical and optical detection of PEF spins using quantum-mechanical selection rules.

### 5.1. Microwave Detection in Microresonators and Superconducting Cavitites

The sensitivity of 'normal' electron spin resonance (ESR) is limited by polarization issues in the same manner as NMR is (see §1.2). The electron spin Zeeman energy is

~$10^3$ times larger than for nuclei, so that thermal electron spin polarization can surpass 50% at realistic conditions (W-band, conventional helium cryostat).

The sensitivity of commercial spectrometers is in the region of $10^9$ spins/$\sqrt{Hz}$; it can be enhanced substantially by using micro-resonators or superconducting microwave cavities. Experiments with N@$C_{60}$ in micro-resonators[100,101] showed a spin sensitivity of roughly $1.8 \times 10^4$ spins/$\sqrt{Hz}$ at room temperature, and calculations extrapolating from this result showed that single-spin sensitivity can be reached at low temperatures with optimized resonator structures. The latest development demonstrated a spin sensitivity of $6.7 \times 10^3$ spins/$\sqrt{Hz}$ for phosphorus in silicon at 10 K,[102] implying that an overnight measurement has indeed single-spin sensitivity. Using superconducting amplifiers and microwave structures, Bienfait *et al.* reported a spin sensitivity of $1.7 \times 10^3$ spins/$\sqrt{Hz}$ for Bi donors implanted in silicon,[103] which is a further improvement by a factor of 4.

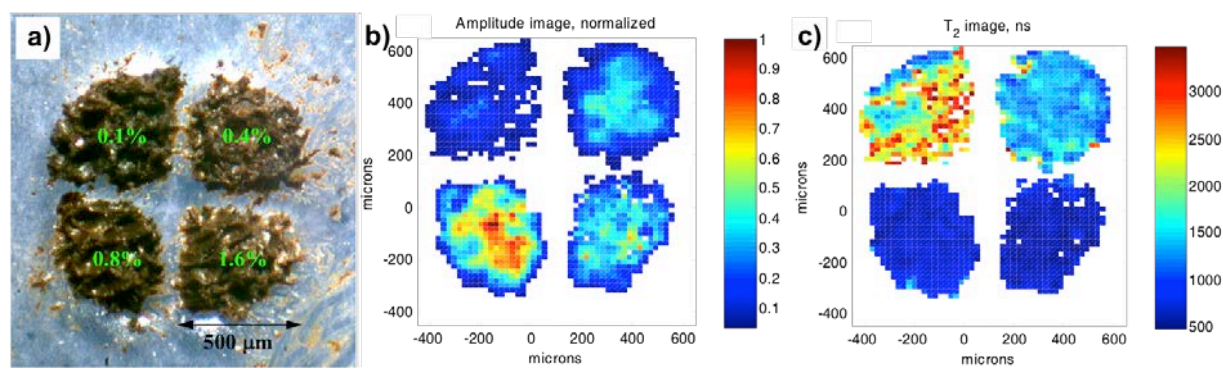

***Fig. 7. High-sensitivity imaging of N@$C_{60}$ in a micro-resonator.*** *ESR images of a heterogeneous sample made of $C_{60}$:N@$C_{60}$ powder with four different N@$C_{60}$ enrichment levels of 0.1%, 0.4%, 0.8%, 1.6%.* ***(a)*** *Optical image showing powder heaps in a 4-quadrant pattern.* ***(b)*** *ESR amplitude image,* ***(c)*** *ESR $T_{2e}$ image. Adapted from Ref.* [100]*, with permission from the Royal Society of Chemistry.*

A major advantage of using non-superconducting microresonators is that high magnetic fields and gradient-imaging methods can be used (see Fig. 7). With proper design, ESR imaging may reach spatial resolutions down to 1 nm. A thorough discussion of this kind of technical issues can be found in the work of Blank *et al.*[100]

Although single-spin ESR sensitivity seems within reach for the near future, the current measurement times are still prohibitive for routine work. Furthermore, inductive spin measurements are 'weak' and not projective, so that this approach will likely not be part of a final quantum computer; rather, the imaging capabilities of the micro-resonator approach may turn out to be an extremely valuable tool in developing extended solid-state spin systems.

### 5.2. Electrical Detection

Electrical detection of spin states in electronic devices is possible using spin selection rules governing carrier transport or carrier recombination (see Fig. 8). In the transport approach, a current is driven through a single endohedral molecule,[104] and Pauli blocking is used to infer the Zeeman state of the coupled system consisting of a charged fullerene cage and the endohedral atom. The second detection approach, known as *electrically detected magnetic resonance* (EDMR), uses spin-dependent recombination of charge carriers in $C_{60}$ thin films[105] or $C_{60}$ microcrystals.[106]

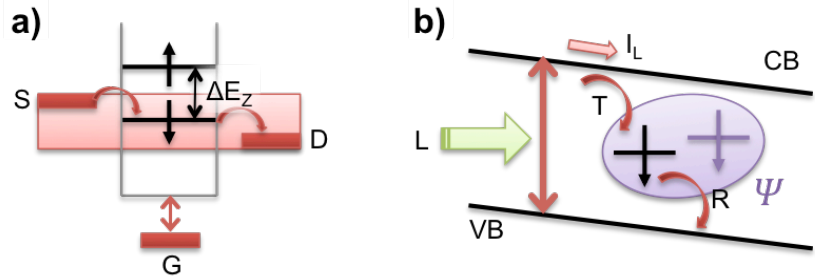

***Fig. 8. Electrical spin detection schemes. (a)*** *Spin-dependent quantum transport through a single molecule. Energy levels (black) defined by the molecular electronic structure — including Zeeman splitting $\Delta E_Z$ — can be shifted with a gate (G) in and out of the bias window (shaded area) defined by source (S) and drain (D) potential difference $V_{SD}$.* ***(b)*** *Spin-dependent photocurrent (SDPC) in a semiconductor. Illumination with light (L) creates carriers in conduction (CB) and valence band (VB), which can be trapped (T) in localized states (black), e.g., as self-localized polarons. The spin state $\Psi$ may prevent or accelerate recombination (R) of the trapped carrier, influencing the macroscopic photocurrent $I_L$ through the device.*

Direct detection of spin levels in a single-molecule-transistor (SMT) device using N@$C_{60}$ was demonstrated by Grose *et al.*[104] and reproduced independently by Roch *et al.*[107] Devices were fabricated by adsorbing almost pure N@$C_{60}$ molecules onto an initially continuous Pt or Au wire, which was deposited on top of an oxidized Al gate electrode. The wire was then broken using electro-migration to form a nanometre-scale gap. Differential conductance measurements at $T \approx 100$ mK as a function of magnetic field $B_0$ showed a transition from a $S = 2$ to a $S = 1$ spin state for the singly charged N@$C_{60}$, i.e. from ferromagnetic to antiferromagnetic coupling between the endohedral $S = 3/2$ and the $C_{60}$ cage $S = 1/2$ spin (see Fig. 9).

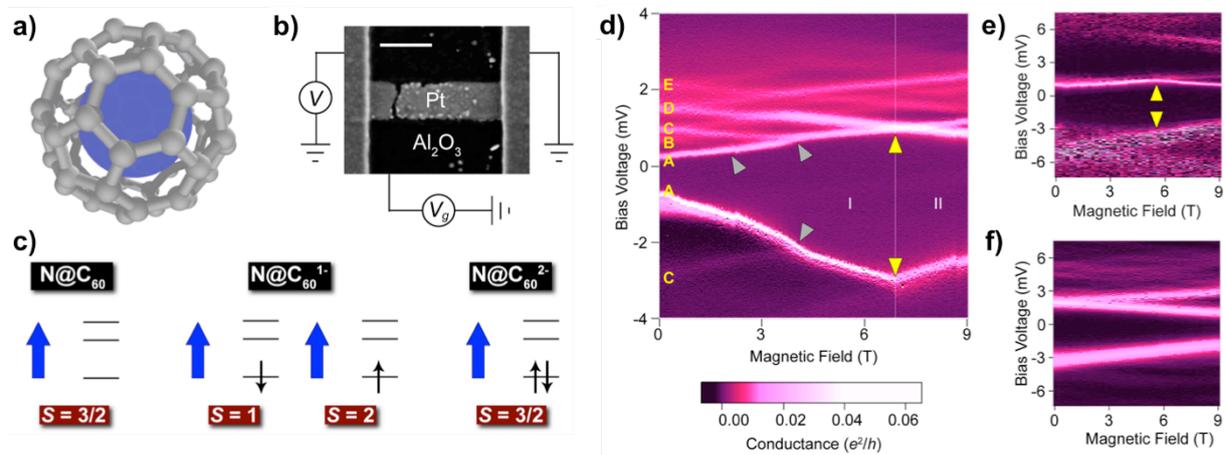

***Fig. 9. Spin states in N@$C_{60}$ single-molecule transistor devices. (a)*** *Sketch of the N@$C_{60}$ molecule.* ***(b)*** *Scanning electron microscopy image of a sample at room temperature following electro-migration (scale bar: 200 nm).* ***(c)*** *Spin states in N@$C_{60}$ and its anions; blue arrows: endohedral spin $S_N$, black arrows: spin $S_e$ of electron(s) in lowest unoccupied orbitals; possible values for total spin $S_q = S_N \pm S_e$ depends on charge state $q = 0,1,2$.* ***(d)*** *Differential conductance $dI/dV$ for N@$C_{60}$ SMT device 1 as a function of bias voltage $V$ and applied magnetic field $B$, at constant gate voltage $V_g = -891$ mV on the positive side of the degeneracy point. In region I, the ground-state transition $(q + 1) \rightarrow q$ corresponds to decreasing spin, $S_{Z,q+1} - S_{Z,q} < 0$, while in region II $S_{Z,q+1} - S_{Z,q} > 0$.* ***(e)*** *Same as (d) for N@$C_{60}$ SMT device 2.* ***(f)*** *Same for a $C_{60}$ SMT device. No change in slope is observed for the ground-state peaks of $C_{60}$. Adapted from Ref. 104.*

Large exchange couplings of several Tesla (>0.1 meV) between the endohedral atom and electrons on the $C_{60}$ cage were determined, but the exact value varied considerably from sample to sample. These variations indicate that the physical situation may be more complex than the simple picture of isotropically coupled endohedral and cage spin. In particular, the anisotropic distribution of charges on the cage, which oscillates between one and two electrons in the experiments, may strongly influence the magnetic coupling. The reported measurements constitute single-molecule quantum state measurements, but they are quasi-static and only report the spin's ground state so far. For use in quantum computation, they should be combined with fast gating[108] and microwave manipulation of the electron spin[109] as has been shown for other solid-state spin qubits, and possibly with spin-filtering contacts.[110] These techniques are already very challenging *per se* and it turns out that reproducible device fabrication is even harder for single-molecular devices than for top-down fabricated qubits such as GaAs quantum dots.

Indirect electrical spin detection via carrier pair recombination is known as EDMR. This technique naturally includes coherent spin control with MW pulses, making it a good platform for quantum computing studies. Pulsed EDMR has been applied to the Kane silicon qubit at low temperatures.[111] Studies on $C_{60}$ thin film devices found spin pair states with lifetimes in the microsecond range even at room temperature,[105] which is particularly attractive for keeping measurement times short. A sensitivity in the range of ~$10^3$ detected elementary charges was reached in these experiments, but EDMR is capable of even higher sensitivity.[112] The results obtained for 'empty' $C_{60}$ were used to predict the EDMR spectrum for dipole-coupled N@$C_{60}$ molecules (see Fig. 10a-b),[105] and several groups tried to measure corresponding spectra.

The experimental realization of an EDMR-based read-out has taken a long time due to the necessary development of 'cold' processing methods compatible with the limited thermal stability of PEFs (see §3.1 and §3.2). Using a table-top cw ESR spectrometer with fast sample-changeover times,[106] we have recently accelerated and improved our sample development. Fig. 10c shows a room-temperature cw ESR spectrum of N@$C_{60}$-doped microcrystals on a grid electrode. The spectral features are rather more complicated than the naïve prediction from 2007, which neglected the anisotropy and distance dependence of dipolar coupling. The observed spectrum is compatible with simulations of an ensemble of strongly dipolar-coupled spin pairs consisting each of one ($S = 3/2, I = 1$) N@$C_{60}$ and one ($S = 1/2, I = 0$) $C_{60}^{\bullet+}$ radical molecule. The pairs are assumed to have a random orientation but a fixed average distance of 1.1–1.2 nm, yielding an average dipolar coupling constant of 30–35 MHz.

These preliminary results are encouraging for further studies using pulsed EDMR at variable temperatures. They need to be extended with pulsed excitation and detection in order to assess coupling strength and relaxation times more quantitatively; the rather narrow EDMR line widths (< 0.2 mT) indicate that the good spin properties of N@$C_{60}$ are retained to some degree in these experiments. It will be interesting to see how far EDMR sensitivity can be pushed by advanced device design. A yet unsolved problem is how to gain control over the dipolar coupling orientation. Carbon nanotube encapsulation of dimers would allow this, but CNT EDMR has yet to be demonstrated.

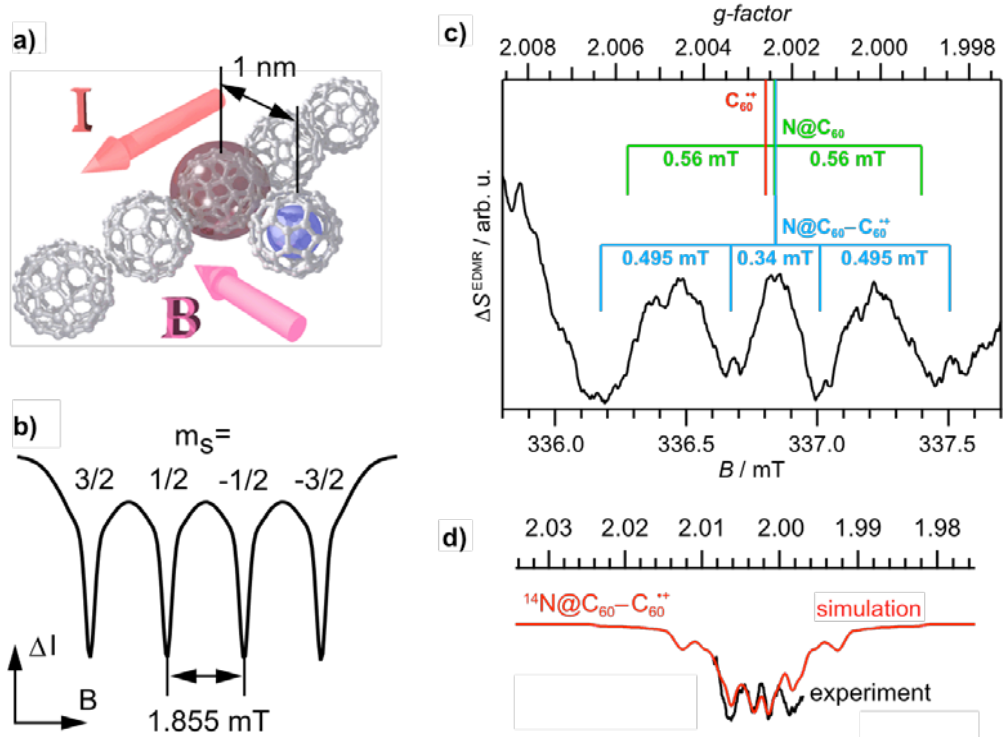

***Fig. 10. EDMR detection of endohedral fullerene spins. (a)*** *Scheme for the EDMR readout of $N@C_{60}$ electron spin states by nearby fullerenes. The localized $C_{60}$ spin involved in spin-dependent transitions, marked by a surrounding sphere, is dipole-coupled to an adjacent $N@C_{60}$ molecule.* ***(b)*** *Simulated integrated cw-EDMR spectrum expected for the $(S(N@C_{60}) = 3/2, S'(C_{60}) = 1/2)$ spin pair with scalar coupling strength $J = 52$ MHz (orientation dependence neglected). Adapted from Ref.* [105]*, with permission from The American Physical Society).* ***(c)*** *Observed integrated cw-EDMR difference spectrum (average of two samples) constructed by subtraction of a broad resonance due to uncoupled cationic $C_{60}^{\bullet+}$ radicals. Line positions of constituent $C_{60}^{\bullet+}$ (measured by EDMR) and $N@C_{60}$ (by ESR) are indicated. The spectrum can be qualitatively simulated with a powder average of dipolar coupled $(S = 3/2, S' = 1/2)$ spin pairs with average $J_D = 30\text{--}35$ MHz, corresponding to a distance $d = 1.1\text{--}1.2$ nm. Adapted from Ref.* [113]*.*

### 5.3. Optical Detection

Optical spin detection uses similar selection rules as electrical detection. In optics, the detection efficiency is boosted by the large photon energy and furthermore provides a way for fast spin polarization. In specific systems with paramagnetic ground states, single-spin sensitivity can be reached at room temperature. These properties have made the nitrogen-vacancy center (NVC) in diamond[19] popular as a highly controllable single-spin quantum bit.[21]

Fullerenes have optical transitions that can be used in a similar way, but usually they have a diamagnetic ground state. The controllable spin thus exists only for several microseconds in a metastable state reached by inter-system crossing and is not particularly suited as a logical quantum bit. Nevertheless, transient spin states can be used for spin state read-out,[114] especially if their creation is accompanied by optical spin polarization. Of particular interest are paramagnetic endohedral fullerenes containing transition metal ions, like $Er_3N@C_{80}$, which have spin-dependent luminescence at low temperature.[6]

Unfortunately, $N@C_{60}$ offers no direct optical access to the spin since the nitrogen atom is extremely well decoupled from the carbon cage, leading to independent energy levels;

the first allowed transition of the nitrogen atom is in the deep ultra-violet (> 10 eV). Therefore, coupling to read-out systems is needed. Since NVCs have the proven potential for detecting weakly coupled electron spins with single-spin sensitivity,[115,116] we have started investigating this system as a read-out.

Spin-spin coupling experiments were performed by depositing enriched N@$C_{60}$:$C_{60}$ mixtures from solution onto diamond surfaces implanted with shallow NVCs. First studies using cw-ODMR remained inconclusive,[117] i.e. the coupling was not large enough for a detectable line splitting. Pulsed ODMR measurements of similar samples (see Fig. 11) showed Rabi oscillations of coupled 'dark' electron spins. It is difficult to quantify the number of coupled spins from these experiments that suffered from a rather low spin coherence time of the NVC readout qubit (< 1 μs).[118] Thus, a full characterization of the dark spins was not possible and coherent coupling control has yet to be demonstrated.

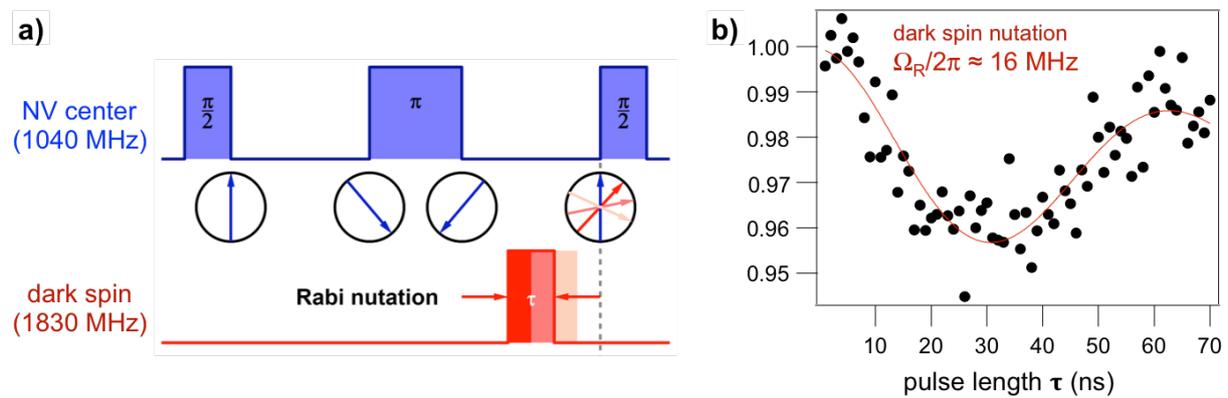

*Fig. 11. ODMR detection of dark spins near a single nitrogen-vacancy center in diamond. (a) Pulse scheme for indirect detection of optically inactive ('dark') spins by a single NVC. An optically detected spin echo is performed on the NVC, while nearby dark spins are driven with microwave pulses. This results in perturbed echo formation and hence diminished luminescence at resonance. (b) Rabi nutation of one or several dark spins, presumed to be N@$C_{60}$, as detected by the modulated echo intensity of a single NVC. Adapted from Ref.* [118].

A current topic of research is hence to increase the spin coherence times of NVCs shallow enough to sense external spins deposited onto the surface. We found that NV centers in nano-diamonds (diameter < 25 nm) are useful for realizing small quantum algorithms at room temperature[119] and for sensing small magnetic fields in their environment.[120] NVCs in nano-diamonds however still show shorter relaxation times than NVCs in bulk diamond. Coherence can be prolonged by using decoupling pulses.[118,121] Recent efforts to optimize NVC for external spin sensing have lead to an intense scrutiny of diamond surface. Since scanning tunneling microscopy (STM) cannot be used with ease on insulating surfaces, we have employed atomic-resolution non-contact-AFM.[122] In conjunction with photoelectron spectroscopy, this has helped us set up protocols to reproducibly prepare ultra-clean diamond surfaces.[123] The latest developments in NVC research show that high-resolution mapping of external spins is possible,[23,121] indicating that the nitrogen-vacancy center can indeed be used as a quantum state detector.

## 6. Summary and Outlook

Ensemble experiments on pnictide endohedral fullerenes have confirmed the potential of this class of molecules for quantum information. In order to progress from this potential to actual devices, a number of challenges have to be overcome.

First and foremost, the synthesis route must be simplified and made efficiently scalable since the scarcity of pure endohedral material has hindered progress in this field considerably. The low overall yield (see §3.1) can be improved at two stages: (a) the ion implantation yield of $3\times10^{-4}$ should be raised at least into the percent range; (b) multi-step HPLC must be avoided by finding better stationary phases, possibly based on molecular recognition.

The second major milestone is the demonstration of a fast and strong read-out at convenient experimental conditions (moderate temperatures, no vacuum, etc.). This may be achieved either via electrical or optical detection. The EDMR route has to be extended down to the single-spin level, which requires developments in device architecture. The most promising approach is optical detection via nitrogen-vacancy centers in diamond or similar detection systems. Recent investigations on the stability of N@$C_{60}$ fullerenes[56] show that the combined optical and thermal load has to be somewhat reduced to enable long-term stability.

With this milestone, spin polarization and large-scale operation in one-dimensional chains will become possible. Such systems can be prepared by the methods indicated in §3, immediately leading to thousands of molecular qubits lined up for experiments. Their investigation will enable a realistic assessment of the ultimate system scaling size for a PEF-based quantum register. Finally, integration of medium-scale quantum registers into hybrid quantum systems[124] will be of interest.

### Acknowledgments

We gratefully acknowledge support by the Deutsche Forschungsgemeinschaft (Heisenberg Program and Priority Program 1601: New Frontiers in Sensitivity for EPR Spectroscopy) and by the VolkswagenStiftung (Initiative: Integration of Molecular Components in Functional Macroscopic Systems).